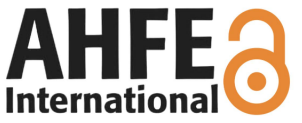

# Human Driver Temperament and the Safety Impact of a C-V2X Denial-of-Service Flooding Attack in Mixed-Autonomy Traffic

**Rasheed Bello[1], Gurcan Comert[2], Varghese Vaidyan[3], Akinbobola Jegede[1], Vijay Bendigeri[4], and Judith Mwakalonge[1]**

[1]South Carolina State University, Orangeburg, SC 29117, USA
[2]North Carolina A&T State University, Greensboro, NC 27411, USA
[3]Dakota State University, Madison, SD 57042, USA
[4]Independent Researcher, Sacramento, CA, USA.

## ABSTRACT

Cooperative and connected automated vehicles (CAVs) rely on Signal Phase and Timing (SPaT) messages to cross signalized intersections; a denial-of-service (DoS) flood that blocks SPaT forces CAVs into a fail-safe mode. Because human-driven vehicles share the intersection, the safety consequence depends not only on the attack and the CAV fail-safe policy, but on how the surrounding human drivers behave. We investigate this human-factors dimension with a coupled OMNeT++/INET (5G NR-V2X) and SUMO microsimulation of a signalized corridor, sweeping CAV market penetration (10–90%), four calibrated driver temperaments (cautious to aggressive) and two standards-based fail-safe policies, a minimal-risk maneuver (MRM) and an adaptive cruise control (ACC) keep-driving fallback, with each attack arm differenced against its policy-matched no-attack baseline. Temperament's effect on the attack is specific and modest rather than a blanket amplification. Aggressive surroundings worsen one metric, the hard-braking a keep-driving fail-safe forces on nearby drivers ($p = 0.03$), rising from near zero to +8 episodes/1000 veh-s. They appear to dampen rear-end conflicts, but only because the flood clears the queues aggressive drivers build, so the gain is in flow, not safety. On the attack's primary signatures, CAV red-light running and crossing conflicts, temperament has no detectable effect. It instead dominates baseline risk, producing a 13- to 18-fold cautious-to-aggressive gradient far larger than the attack itself, which acts through a channel already congested by CAV adoption. Human driver populations determine how dangerous the intersection is but do not systematically amplify this attack, so fail-safe design cannot assume a cautious test population bounds the risk.

## INTRODUCTION

Signalized intersections concentrate crash risk, and the connected-and-automated-vehicle (CAV) program treats them as the place where vehicle-to-everything (V2X) communication can retire human error. A CAV approaching a signal can read the lamp with onboard cameras, but the Signal Phase and Timing (SPaT) message broadcast over cellular vehicle-to-everything (C-V2X) carries the time-to-change a camera cannot see and survives occlusion or loss of line of sight. A CAV that relies on SPaT for the stop-or-go decision, the case we model, must have that message at the instant of decision. That dependency is the attack surface. If an adversary blocks that message, the vehicle loses the one input telling it the signal is red.

C-V2X is a documented denial-of-service (DoS) target. Trkulja et al. (2020) show that C-V2X Mode 4 scheduling can be abused to induce collisions, while Twardokus and Rahbari (2023) demonstrate stealthy attacks that cut a target's packet delivery ratio (PDR) by 90% within seconds on a hardware testbed. Tine et al. (2025) close the gap to practice: protocol-compliant flooding of a commercial on-board unit cuts PDR by 87% and suppresses forward-collision warnings entirely. The attack is real, cheap, and effective. What remains uninvestigated is what happens on the road once this attack succeeds.

A CAV does not fail in isolation. For decades it will share intersections with human-driven vehicles and SPaT loss does not remove those humans — it changes the automated vehicle's behavior in their midst. Denied its signal phase, a CAV falls back to a standards-defined safe state: a minimal-risk maneuver that brings it to a controlled stop (SAE, 2021; UNECE, 2021), or a graceful degradation from cooperative to conventional adaptive cruise control that keeps it moving (Ploeg et al., 2013). Either choice propagates into the surrounding traffic, and the drivers who inherit it are not uniform. They range from cautious drivers who yield generous gaps to aggressive drivers who accept short gaps, run late-phase signals, and refuse to cede right-of-way (He et al., 2025; Yao et al., 2024). The safety outcome of the attack is therefore not a property of the network alone; it is set at the point where a blinded CAV meets a heterogeneous human network.

Existing work does not occupy that point: the security literature stops at the radio (Tine et al., 2025; Trkulja et al., 2020; Twardokus & Rahbari, 2023). the control literature treats surrounding traffic as a disturbance rather than a moderator (Ploeg, et al., 2015; Wang, et al., 2023). and human-factors studies of driver temperament describe ordinary mixed traffic, without an attacker in the loop (He et al., 2025; Li et al., 2023; Yao et al., 2024). No study measures, at population scale, how the surrounding human driving population shapes the safety consequence of a C-V2X attack once a CAV enters its fail-safe mode.

We close this gap with a coupled network-traffic microsimulation of a signalized corridor under a SPaT-denial flood, and we make three contributions. First, we build a measurement framework. We couple a network-layer attack model to a microscopic traffic model and drive it with a matched-differencing design across a penetration-by-temperament-by-policy sweep, which isolates the attack's marginal safety effect from the confounds of demand, geometry, and CAV penetration. Second, and centrally, we answer an open human-factors question: do surrounding drivers' temperaments damp or worsen the attack? We show that temperament sets the baseline level of intersection risk but does not amplify the attack in any significant manner. Third, we run the full sweep under two CAV fail-safe policies, attributing the outcome to the vehicle's response strategy and testing how it interacts with temperament. Together, these contributions reframe C-V2X denial of service from a network-security statistic into a human-factors safety problem.

## RELATED WORK

### Attacks on C-V2X and fail-safe degradation

Three literatures bound this problem without meeting at it. Attacks on C-V2X exploit its decentralized scheduling: Trkulja et al. (2020) report that at high vehicle density even a crude oblivious attacker is nearly as effective as a sophisticated one, because the channel is already contended; Twardokus and Rahbari (2023) collapse a target's PDR by 90% at the physical and MAC layers; and Tine et al. (2025) achieve the same by saturating the receiver host rather than the radio. All three stop at the radio and none carries the loss forward into vehicle motion, which is where our analysis begins; our flood, which collapses SPaT delivery to roughly 0.2 at high adoption, is a magnitude-consistent analogue of their validated results. On the vehicle side, CACC collapses to conventional ACC under packet loss and loses string stability (Milanés & Shladover, 2014; Ploeg et al., 2013), a degradation Tu et al. (2019) find measurably worsens longitudinal safety, while the alternative is the standards-defined controlled stop (SAE International, 2021; UNECE, 2021). Our two fail-safe arms instantiate this dichotomy; prior work evaluates such policies on the automated vehicle's own dynamics rather than on the humans downstream.

### Driver temperament in mixed traffic

Driver heterogeneity is equally well established: He et al. (2025) cluster field trajectories into conservative, normal, and aggressive styles in which aggressive drivers accelerate to close the gaps conservative drivers yield. Li et al. (2023) and Yao et al. (2024) show that car-following heterogeneity produces a safety gradient. We adopt this typology as the moderator variable, extending the three-cluster field taxonomy to a four-level sweep, and, unlike these studies of ordinary traffic, place it under attack.

### Surrogate safety measurement

Because simulated crashes are rare and unreliable, safety is assessed through surrogate safety measures (SSMs) computed from vehicle trajectories. USDOT FHWA (2003, 2019) established time-to-collision (TTC) and post-encroachment time (PET) as the canonical conflict measures and the microsimulation calibration discipline behind them. Das et al. (2023) name the open problem we take seriously: threshold selection is unsettled and the conflict-to-crash ratio is not fixed. This motivates our two safeguards: reporting effects as deltas against a policy-matched baseline and retaining only SSMs that behave monotonically.

## METHOD

### Coupled simulation platform

We couple a network simulator and a traffic micro-simulator so that communication and motion evolve together (Figure 1). The network side is OMNeT++ with the INET framework (Varga, 2010), carrying a NS-3 learned 5G NR-V2X PC5 surrogate model (Bello et al., 2026). The traffic side is SUMO (Lopez et al., 2018), running microscopic car-following and junction models on a calibrated network of the Chestnut Street corridor, SC. The two exchange state every time step over the Traffic Control Interface (TraCI), following the bidirectional coupling of Sommer et al. (2011): SUMO advances vehicle positions, OMNeT++ resolves which SPaT and basic safety messages are delivered, and delivery outcomes feed back into each CAV's control decision. Roadside units at each signalized junction broadcast SPaT at 10 Hz; CAVs use it to time their approach.

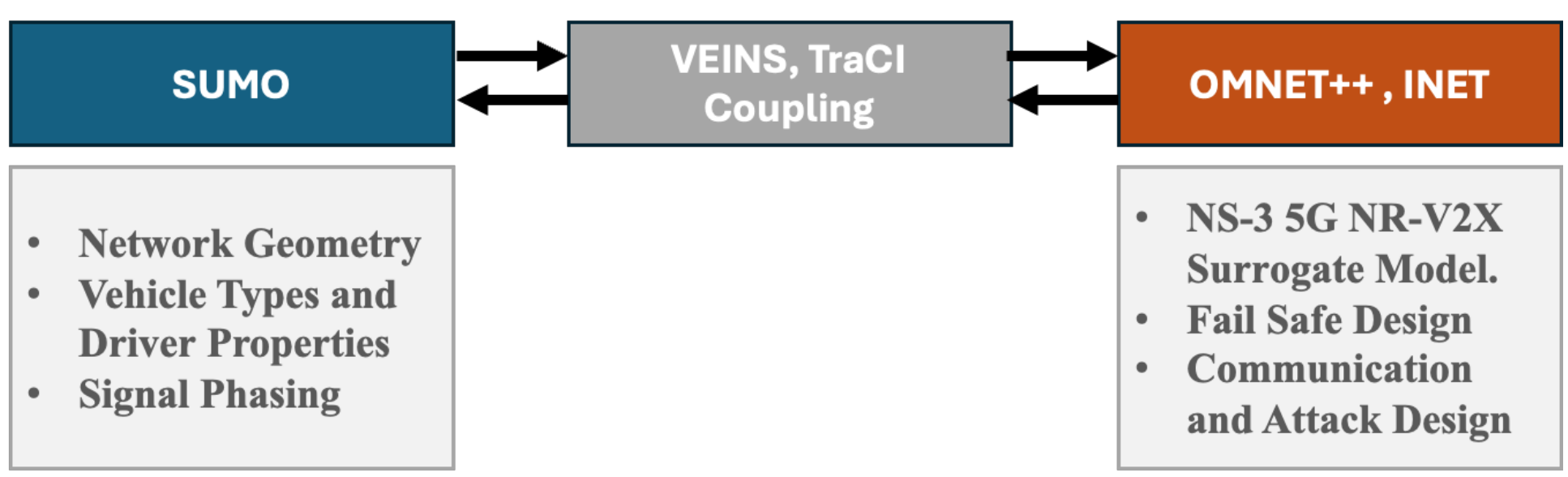

*Figure 1: Coupled simulation setup on SUMO, VEINS, OMNET++/INET*

### Communication model

SPaT and safety messages travel over a 5G NR PC5 link that we represent with a surrogate learned from ns-3 5G-LENA and release as code (Bello et al., 2026). The application carries two SAE J2735-style messages over UDP on this link: 256-byte SPaT broadcast by roadside units at 10 Hz, which CAVs consume to time their approach, and 256-byte basic safety messages (BSM) broadcast by vehicles at 10 Hz. Each message is delivered with a probability that combines a learned radio stage with a processing limit at the on-board unit,

$$\boldsymbol{PDR}(\boldsymbol{d},\boldsymbol{n}) = p_{rf}(d,n)(1 - p_{obu}) \quad (1)$$

where $\boldsymbol{d}$ is the transmitter-to-receiver distance and $\boldsymbol{n}$ the local CAV density. The radio term $\boldsymbol{p_{rf}}$ is a cascade fit to ns-3: it falls with distance through the 3GPP TR 37.885 V2V-Urban path loss and SINR, and with density through a half-duplex loss, a semi-persistent-scheduling collision, and a decode waterfall with a capture branch that lets near, high-SINR links survive collisions. Each stage is a logistic function of $\boldsymbol{d}$ or $\boldsymbol{n}$ with coefficients fit by logistic regression, listed in Table 1. This per-mechanism decomposition follows the analytical C-V2X model of Gonzalez-Martin et al., (2019); Bello et al., (2026) extend that line by learning the stages rather than deriving them. SPaT reliability at a junction is then the mean of independent per-message deliveries over the analysis window. Keeping delivery both distance- and load-dependent lets channel self-congestion and the flood both register while making the full penetration, temperament, and policy sweep tractable.

*Table 1: CV2X surrogate model summary*

| Component | Configuration |
| --- | --- |
| SINR (PHY) | $\boldsymbol{\gamma(d) = P_{tx} - PL(d) - N_0}$<br>$P_{tx}$= 23 dBm, $N_0$= −94.28 dBm, $f_c$ = 5.9 GHz; PL: TR 37.885 V2V-Urban |
| Half-duplex | $h(n) = \sigma(\eta_0 + \eta_1 n + \eta_2 n^2)$<br>η = (−2.4489, 9.94e−3, −7.7e−5) |
| Decode, clean | $g^{(nc)} = \sigma(\theta_0 + \theta_1 \gamma + \theta_2 \gamma^2 + \theta_3 \gamma + \theta_4 \gamma n)$<br>$\theta^{(nc)}$= (0.00359, 0.05230, 0.00389, 0.08711, −0.00325) |
| Decode, capture | $g^{(cc)} = same\ expression\ as\ g^{(nc)}$<br>$\theta^{(cc)}$= (−3.16246, 0.05439, 9.1e−4, −0.02843, 1.6e−4) |
| Radio cascade | $P_{rf} = (1 - h)[(1 - c)g^{(nc)} + c\, g^{(cc)}$ |
| Flood factor | $q_a(r, n) = \sigma(\omega_0 + \omega_1 \log_{10} r + \omega_2 \log_{10}(n+1) + \omega_3 \log_{10} r \log_{10}(n+1))$ |
| Collision, base (MAC) | $c_0(n) = \sigma(\kappa_0 + \kappa_1 n + \kappa_2 n^2)$<br>κ = (−2.81848, 0.17125, −9.273e−4 |
| Collision under Flood (MAC) | $c(n) = 1 - (1 - c_0(n)) \prod_a (1 - q_a)$ |
| OBU overload | $p_{obu} = min\left(1, smax\left(0, \frac{L - C}{L}\right)\right)$<br>$C = C_{obu}$ ; W ; $C_{obu}$ = 600 pkt/s, |

## Threat model and attack

The adversary is a single stationary flooder parked at the target junction, transmitting junk traffic at 1000 packets per second. This exceeds the modeled 600-packet-per-second processing capacity of an on-board unit, so the flood starves the receiver of the budget it needs to decode legitimate SPaT, the receiver-path saturation Tine et al. (2025) observe in hardware, rather than raw radio jamming. We deliberately use a blunt, non-intelligent attacker: because the channel is already contended at high adoption (Trkulja et al., 2020), a simple flood suffices, and the result is a lower bound on what a sophisticated adversary could achieve. The attack is active over a 100 s window (t = 100–200 s); safety is measured within a 150 m analysis disc centered on the target junction.

## Fail-safe policies

A CAV's SPaT stream goes stale after 0.5s of no valid message, after which it commits to one of two fail-safe policies after a 1s fallback handover. The minimal-risk maneuver (MRM) executes a controlled, dilemma-zone-aware stop at a comfortable 3.0 m/s² deceleration, within the UNECE, (2021) envelope and consistent with the minimal-risk condition of SAE, (2021), holding upstream of the stop line until fresh SPaT returns. The ACC fallback is the graceful CACC-to-ACC degradation of Ploeg et al. (2013): the CAV drops cooperative control and keeps driving on its own sensors, which, blind to the red phase, allows it to proceed. The two policies differ only in the SPaT-loss branch. Because baseline SPaT delivery is imperfect (PDR ≈ 0.30), even a no-attack CAV occasionally loses SPaT long enough to enter that branch, so base-MRM and base-ACC need not be identical; we therefore simulate a separate no-attack baseline for each policy and difference every attack arm against its own policy- and seed-matched baseline.

### Driver temperaments

Human drivers follow the Intelligent Driver Model while CAVs rely on the CACC and ACC controllers introduced earlier. We define four driver-temperament profiles that span cautious to aggressive. Unlike He et al. (2025), who identify three driver styles by clustering trajectories at a single signal-free freeway merge, we specify our profiles directly and anchor each in measurable behavior: the gaps and headways a driver accepts, how readily it forces right-of-way, and how far past a signal change it proceeds. This also captures signalized-junction behaviors that a freeway-only typology cannot provide.
Each simulation run contains only one temperament profile and we sweep profiles across runs instead of mixing them within a single run. Aggressiveness is expressed consistently across eleven behavioral parameters. From cautious to aggressive, the profile means move monotonically: desired-speed factor 0.90 to 1.20, impatience 0.05 to 0.96, minimum gap 3.01 to 1.28 m, headway 2.00 to 0.81 s, drive-after-red time 0.02 to 0.90 s, and ignore-foe probability 0.00 to 0.40, with lane changing growing more assertive and less cooperative over the same range. That monotonicity is what lets us treat temperament as a controlled, ordinal experimental factor rather than a subjective style label. To avoid identical behavior within a profile, each parameter is drawn per driver from a Gaussian distribution centered on the profile mean and clamped to that profile's band. Junction-behavior noise is sampled likewise, so no two human drivers behave exactly alike.

### Experimental design

The design crosses CAV market penetration (10, 20, 40, 60, 80, 90%), driver temperament (cautious, normal, assertive, aggressive), fail-safe policy (MRM or ACC), and attacker presence, with ten random seeds, for 954 completed runs of a nominal 960. Roadside units broadcast over a 1000 m range and vehicle-to-vehicle links over 300 m. Critically, both no-attack baselines are simulated (base-MRM and base-ACC), so that each attacked run is differenced against a baseline sharing its policy, penetration, temperament, and seed, forming a matched 2 × 2 (baseline vs attack) × (MRM vs ACC) within every cell. This matching is the backbone of the analysis: it removes demand, geometry, and temperament as confounds, leaving the attack as the only thing that changed.

### Safety metrics and estimation

Safety is quantified with surrogate safety measures computed from trajectories ( USDOT FHWA, 2003; Das et al., 2023). Rear-end risk is a time-to-collision (TTC) conflict: for a moving follower closing on its leader, TTC is the bumper-to-bumper gap over the closing speed; a conflict is recorded when $0 < \text{TTC} < 1.5$ s at a closing speed above 2 m/s.

Crossing risk is a post-encroachment-time (PET) conflict: when two different vehicles on crossing headings (heading difference $\geq 45°$) occupy the same 3 m cell, PET is the gap between the first leaving and the second entering; a crossing conflict is recorded when $\text{PET} < 2$ s.

A hard-braking event is a braking episode whose peak deceleration reaches 4.0 m/s$^2$, and CAV red-light running counts each crossing of the stop line on a red phase, reported per 100 CAVs. Two estimation steps make these counts comparable. First, because the number of vehicles inside the analysis disc varies across runs, raw counts are normalized by exposure, giving for each metric a rate equal to a reporting scale k times its raw count divided by exposure E, where $\boldsymbol{C_m}$ is the raw count of metric $\boldsymbol{m}$, $\boldsymbol{E}$ is exposure in vehicle-seconds (the summed dwell time $\boldsymbol{t_v}$ of every vehicle $\boldsymbol{v}$ in the analysis window), and $\boldsymbol{k}$ is a reporting scale ($\boldsymbol{k}$ = 1000 for rear-end conflicts per 1000 vehicle-seconds; red-running per 100 CAVs). Second, to isolate the attack from every confound, each attacked run is differenced against its matched no-attack baseline sharing policy **p**, temperament $\boldsymbol{\tau}$, penetration $\boldsymbol{\rho}$, and seed $\boldsymbol{s}$; the isolated effect is the attacked minus the baseline value of the metric within that cell.

The difference is the marginal effect of the attack, free of baseline temperament and demand. Whether temperament changes the size of that effect for a given safety metric, rather than merely its baseline level, is tested by regressing the isolated effect on aggressiveness, policy, and their interaction, with a random intercept over penetration and seed. The same model is fit to each safety metric,

$$\Delta_m = \beta_0 + \beta_1\alpha_\tau + \beta_2 x_p + \beta_3\left(\alpha_\tau x_p\right) + \varphi_{\rho s} + \varepsilon \tag{2}$$

The isolated effect $\boldsymbol{\Delta_m}$ is that of safety metric m, the aggressiveness score $\boldsymbol{\alpha_\tau}$ runs 1 to 4, the policy indicator $\boldsymbol{x_p}$ is 1 for ACC and 0 for MRM, and $\varphi_{\rho s} \sim N(0, \sigma^2)$ is the random intercept. The interaction coefficient $\boldsymbol{\beta_3}$ is the quantity of interest: a non-zero $\boldsymbol{\beta_3}$ means temperament amplifies or damps the attack's effect on that metric. The null hypothesis is $\boldsymbol{\beta_3} = 0$.

## RESULTS AND DISCUSSION

### Adoption-limited channel

Two things deny SPaT, and only one of them is the attacker (Figure 2). With no adversary present, baseline SPaT delivery at the intersection falls from 0.61 at 10% penetration to 0.26 at 90%, a 57% loss driven purely by CAVs contending for the shared channel: offered transmissions grow roughly eightfold across the adoption range while successful deliveries saturate, so the delivery ratio erodes as more vehicles connect. The flood then drives what remains down to about 0.18 at high penetration. The attacker does not confront a healthy channel, it rides on one that adoption has already degraded, consistent with the density dependence Trkulja et al., (2020) report and the scaling limits Shah et al., (2024) document. SPaT reliability is adoption-limited before any attack, which is inherently a deployment concern and the backdrop against which the attack must be read.

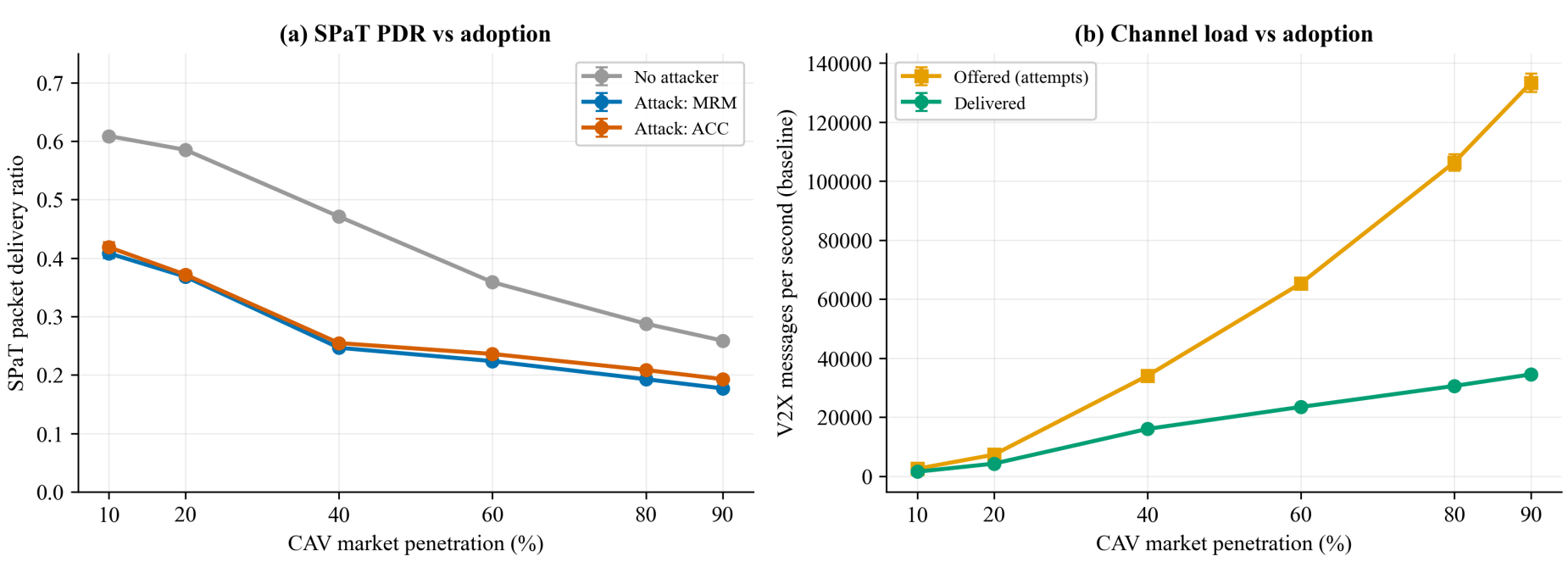

*Figure 2: SPaT delivery versus CAV penetration, with and without the attack.*

## Automated red-light running

The direct consequence of SPaT denial is that a blinded CAV runs the red signal (Figure 3), and the two sources of denial act through one common channel. Pooling every run, a CAV's probability of running a red is set by the delivery ratio it experiences and climbs steeply as delivery falls (Spearman, $r_s = 0.55$, $p < 10^{-74}$). At a given delivery ratio, a self-congested baseline run and an attacked run produce statistically the same red-running, so the source of the loss does not matter, only its size. This is why the attacker is crowded out at scale: once adoption has pushed the baseline onto the low, flat end of that curve, the flood's further drop moves the outcome little. The attack's own marginal effect is therefore policy dependent. Against the matched baseline it raises red-running 1.7 times under ACC ($p < 10^{-7}$; Table 2), while under MRM the increase is not significant, because the ACC fallback keeps a blinded CAV rolling through the intersection whereas the MRM stop holds it upstream of the stop line, its residual violations being dilemma-zone cases where a controlled stop is infeasible.

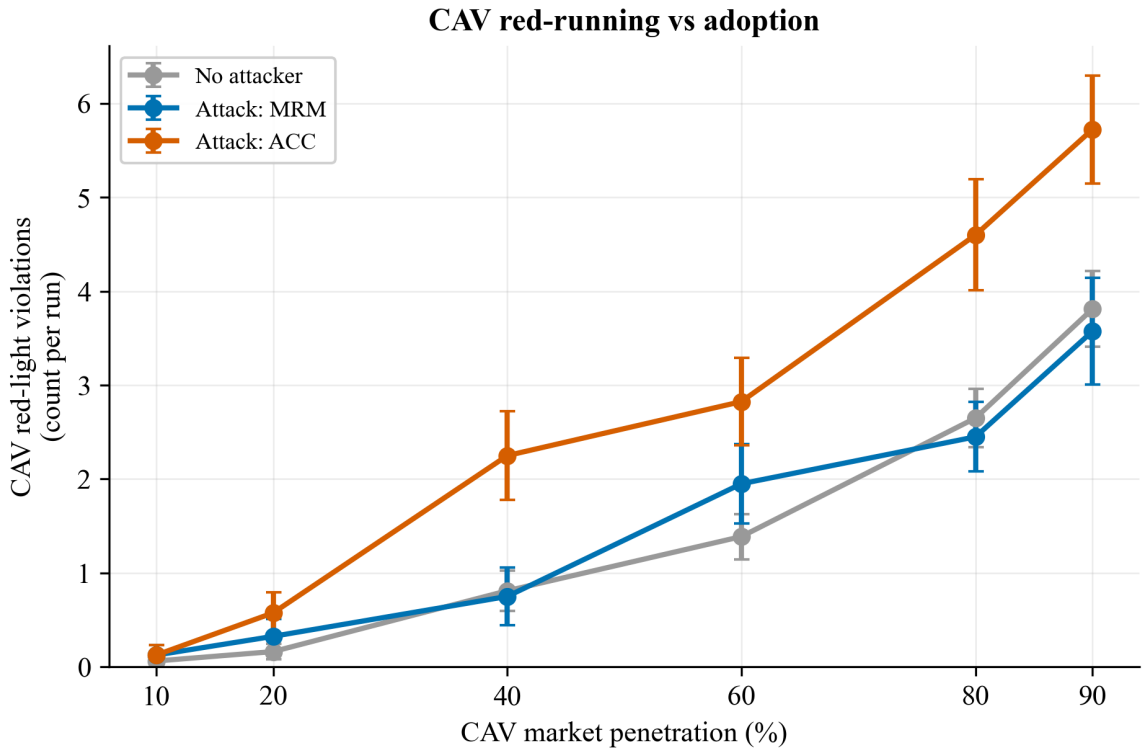

*Figure 3: CAV red-light running versus CAV penetration, by fail-safe policy.*

## Does temperament damp or worsen the attack?

Figure 4 summarizes the result. Testing the attack's isolated effect against aggressiveness, metric by metric, gives three outcomes. In the first, temperament worsens the attack: under the keep-driving fail-safe the flood adds human hard-braking, and that addition grows with aggressiveness, from near zero for cautious surroundings to about +8 episodes per 1000 vehicle-seconds for aggressive ones (interaction $p = 0.03$). Because the blinded CAV keeps rolling through the red phase, a close-following aggressive driver must brake hard where a cautious one would already have left a gap.

In the second outcome, temperament appears to damp the attack, but the appearance is a traffic-flow effect rather than a safety gain. The flood lowers rear-end conflicts most for aggressive populations (interaction $p < 0.001$), not because those drivers become safer but because the flood dissolves the stop-and-go congestion they create: blinded CAVs stop less at signals, the queue clears, and the queue-bound rear-end conflicts clear with it. In the third outcome, temperament has no recordable effect at all. On the attack's primary signatures, CAV red-light running and crossing conflicts, the interaction with aggressiveness is indistinguishable from zero ($p > 0.4$), so the flood acts the same whether the human cars are cautious or aggressive. Temperament thus changes the flood's consequence in one narrow, policy-specific way, leaves the compliance failure untouched, and elsewhere only reshapes the congestion the flood relieves._A study reading only these conflict surrogates would conclude, wrongly, that the flood made the intersection safer, which is why we report red-running directly (Das et al., 2023).

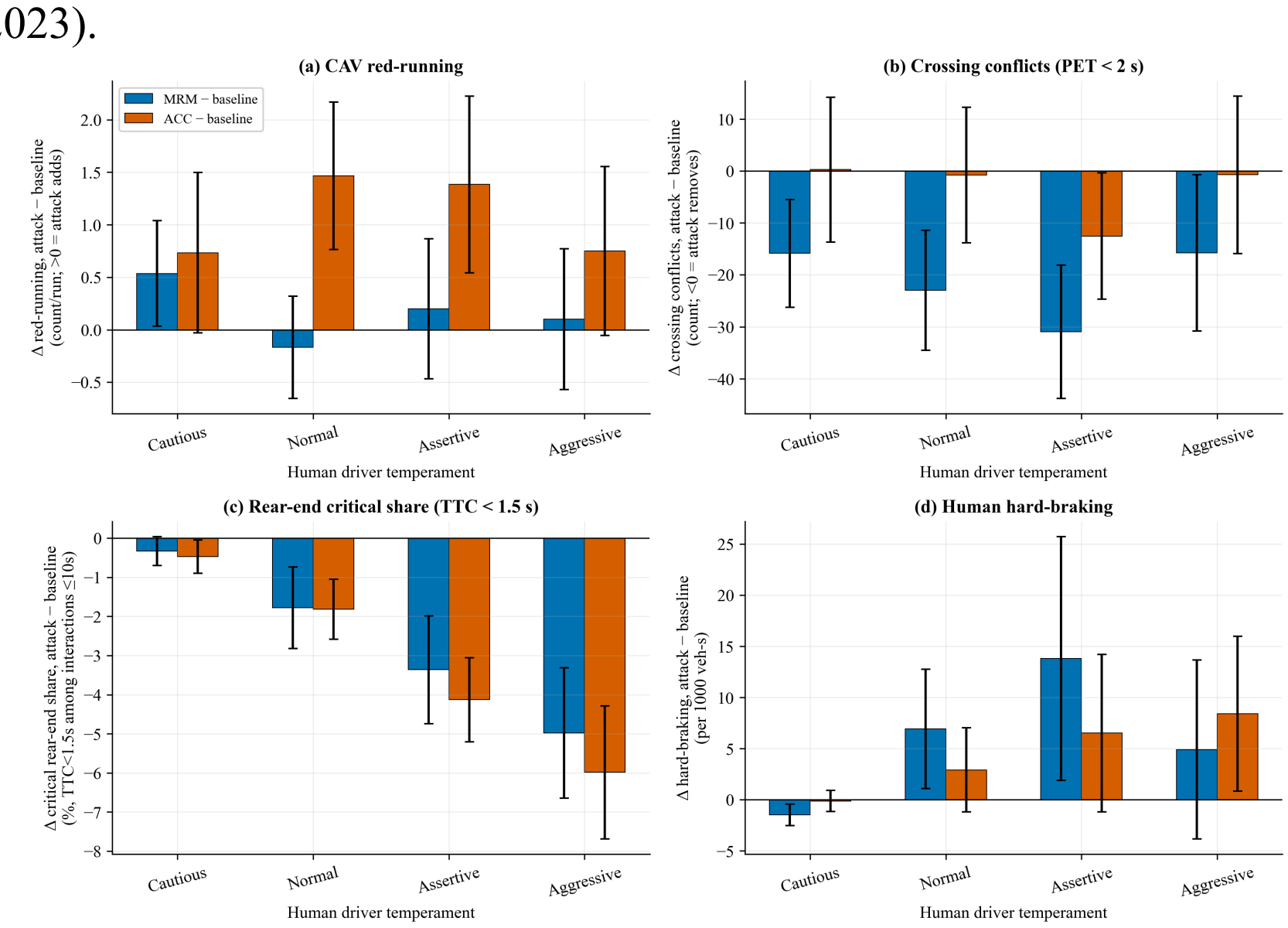

*Figure 4: Isolated attack impact by driver temperament across four safety metrics.*

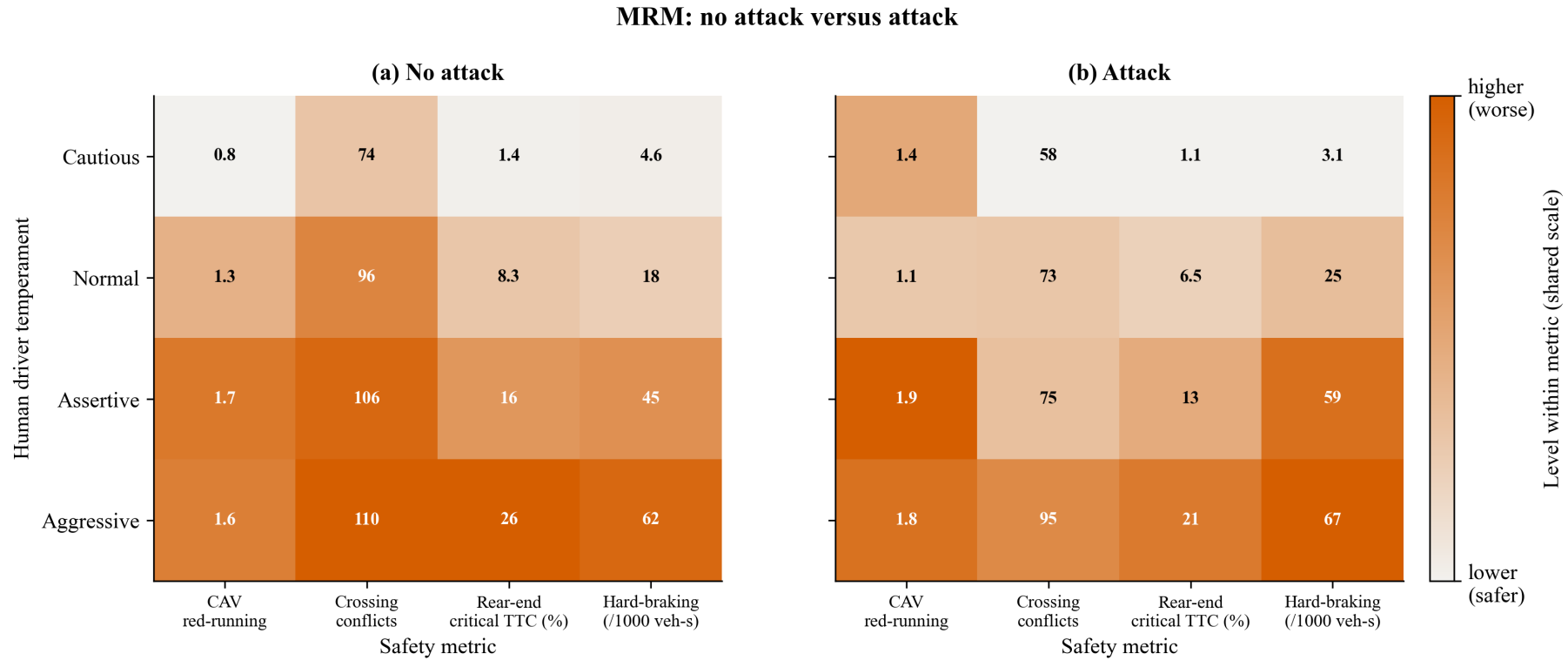

*Figure 5: Safety levels under the MRM fail-safe by driver temperament and metric*

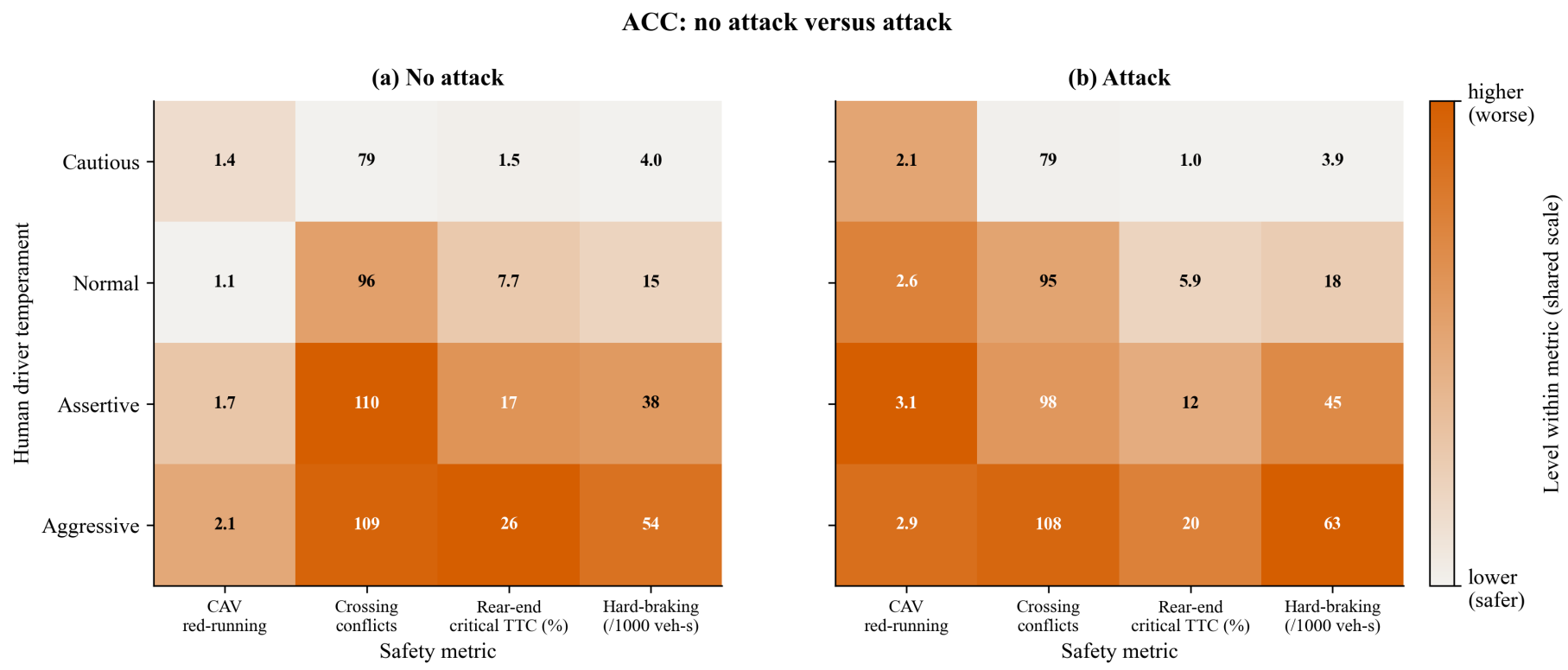

*Figure 6: Safety levels under the ACC fail-safe by driver temperament and metric*

## Temperament dominates the outcome

The two threads of the analysis converge on one place: the human population, not the attacker, governs intersection safety. The policy scorecards show it directly (Figures 5 and 6), since in both the no-attack and attack panels the metrics darken from cautious to aggressive, so temperament sets the level while the attack barely shifts it. That gradient is an order-of-magnitude effect, 18-fold on the critical rear-end share and 13-fold on hard-braking, it is already present with no attacker, and it dwarfs both the flood and the fail-safe policy. The attacker's own footprint is small and, on the metrics that flow from signal compliance, is dominated by self-congestion: raising penetration from low to high adds several times more red-running and crossing exposure at baseline than the flood adds on top (Figure 3), because both act on the CAV through the same loss of SPaT delivery. For practice this cuts two ways: a CAV fail-safe cannot be validated against an average driver, since the one behavior the population amplifies, hard-braking under the keep-driving fallback, is invisible to a cautious test fleet, and because self-congestion is the dominant source of SPaT loss at scale, channel-load control is a first-order safety requirement in its own right.

*Table 2: Baseline correlation of each safety metric with driver temperament, and the matched effect of the attack, by fail-safe policy*

| **Safety metric** | **Baseline r (MRM)** | **Baseline r (ACC)** | **Attack Δ (MRM)** | **Attack Δ (ACC)** |
|---|---|---|---|---|
| CAV red-running | 0.19 ($p < 10^{-3}$) | 0.17 ($p < 10^{-3}$) | +12% (n.s.) | +68% ($p < 10^{-7}$) |
| Crossing conflicts (PET < 2 s) | 0.33 ($p < 10^{-7}$) | 0.29 ($p < 10^{-6}$) | −22% ($p < 10^{-9}$) | −3% (n.s.) |
| Rear-end critical share (TTC < 1.5 s) | 0.95 ($p < 10^{-127}$) | 0.96 ($p < 10^{-130}$) | −2.7 pp ($p < 0.01$) | −3.1 pp ($p < 0.01$) |
| Hard-braking (per 1000 veh-s) | 0.88 ($p < 10^{-78}$) | 0.89 ($p < 10^{-83}$) | +18% ($p = 0.05$) | +16% (n.s.) |

## CONCLUSION

The paper asked whether human temperament damps or worsens a denial-of-service flood that silences SPaT, and we have uncovered that it does both in narrow ways and neither in general. Human temperament dominates how dangerous a signalized intersection is: the gap between cautious and aggressive surroundings dwarfs the flood itself. But it does not systematically amplify the attack. Aggressive populations worsen only the hard braking of nearby human-driven vehicles; they appear to soften the attack on rear-end conflicts only because it relieves the congestion they create; and on the attack's defining signature, automated red-light running, the flood is indifferent to who is driving the human cars. These conclusions carry four caveats. The channel is an analytical surrogate whose absolute delivery ratios need field validation, and the attacker is a single stationary flooder. The temperaments are ordinal profiles we specify directly from a field-derived taxonomy, not human-in-the-loop drivers. The scope is narrow in two ways: automated red-running is rare, so the crossing-conflict null holds only at the tested demand, and temperament's amplification of the attack is confined to hard-braking under the ACC fallback. Future work will validate the channel against field measurements, extend to more geometries and demand levels, and replace parameter-set temperaments with driver-calibrated behavior. While DoS flooding attack appears to be temperament blind for the core safety metric, the findings of this study suggests that fail-safe design must be checked across the full range of human driving behavior for general SPaT Loss, or potentially more damaging attacks.

## ACKNOWLEDGMENT

The authors acknowledge the support and feedback of advisors, colleagues, and reviewers at South Carolina State University and its collaborating institutions. The authors used Anthropic (2026), Opus4.8 for C++ code implementation and audits.